\documentclass[preprint,12pt,authoryear]{elsarticle}

\usepackage{epsfig}

\usepackage{amssymb}
\usepackage{amsthm}

\usepackage{lineno}

\journal{Nuclear Physics A}

\begin{document}
	
	\begin{frontmatter}
		
		
		
		\title{Commissioning and first performances of the ALICE MID RPCs}
		
		
		\author{Livia Terlizzi on behalf of the ALICE Collaboration}

		\affiliation{organization={Università and INFN Torino},
			addressline={Via Pietro Giuria, 1}, 
			city={Torino},
			postcode={10125}, 
			state={Italia}}
		
		\begin{abstract}
			ALICE (A Large Ion Collider Experiment) at the CERN Large Hadron Collider (LHC) is designed to study p-p and Pb-Pb collisions at ultra-relativistic energies. ALICE is equipped with a Muon Spectrometer (MS) to study the heavy charmonia in p-p and heavy ion collisions via their muonic decay. At first, in the LHC Run 1 and 2 the selection of interesting events for muon physics in the MS was performed with a dedicated Muon Trigger system based on Resistive Plate Chambers (RPCs) operated in maxi-avalanche mode. During the Long Shutdown 2 (LS2) of LHC ALICE underwent a major upgrade of its apparatus: since Run 3 (started in July 2022), in order to fully profit from the increased luminosity of Pb-Pb collisions (from 20 kHz in Run 2 to 50 kHz in Run 3), the ALICE experiment is running in continuous readout (triggerless) mode and the Muon Trigger became the Muon IDentifier (MID).
			In order to reduce the RPC ageing 
			and to increase the rate capability, it was decided to use a new front-end electronics FEERIC with a pre-amplification stage to minimize the charge released per hit inside the gas gap. 
			A description of the MID upgrades, together with the results and performances of the RPCs from the commissioning, is presented in this talk.
			
		\end{abstract}
		
		
		
		\begin{keyword}
			RPC, MID, ALICE
			
			
		\end{keyword}
		
	\end{frontmatter}
	
	
	\section{The ALICE Experiment}
	\label{ALICE_MS}
	
	The main goal of ALICE [1][2] is to assess the properties of Quark Gluon Plasma (QGP), a state of matter reached in extreme conditions of temperature and energy density, where quarks and gluons are de-confined [3][4]. One of the main observables used to study the QGP is the production of heavy quarkonia (i.e. bound states of a heavy quark and the corresponding anti-quark) in Pb-Pb collisions. In order to detect quarkonia via their di-muon decays, ALICE is equipped with a forward Muon Spectrometer [5] covering 
	pseudorapidity range 2.5 $< \eta <$ 4. 
	In order to improve the resolution of the secondary vertex and to allow new physics measurements, a silicon-based detector, the Muon Forward Tracker [6], has been installed in front of the MS 
	during the LS2. It is constituted by ten planes of ALPIDE Silicon pixel sensors [7] 
	and it has a overall spatial resolution better than 5$\mu$m. 
	After the MFT, there is a front absorber to stop most hadrons emitted in the MS acceptance in order to reduce the particle flux on the Muon Chambers (MCH). 
	The MCH is a tracking system made of 5 stations of 2 planes of Cathode Pad and Cathode Strip Chambers, with a spatial resolution of 100$\mu$m. The MS has a dipole magnet which 
	provides an horizontal field of 3Tm, perpendicular to the beam axis. 
	Then there is a 1.2 m thick iron wall to filter the residual background of hadrons and finally the Muon IDentifier (MID) [9], 
	described in the next section. \\
	

	\section{The ALICE Muon Identifier (MID)}
	\label{MID_subsec}

\begin{figure}[htb]
		\centering
		\includegraphics[width=8.5cm,height=3.5cm]{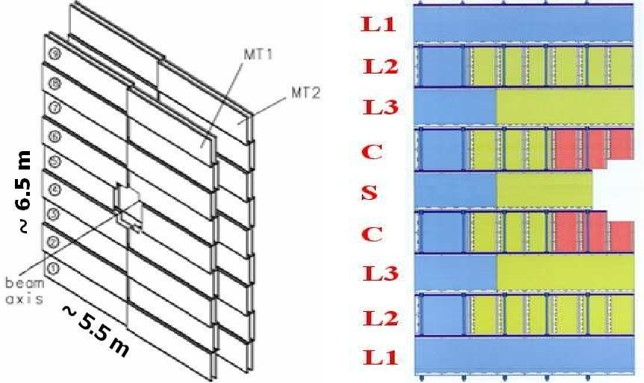}
		\caption{Left: schematic view of the MID. Right: composition of an half plane. The different colors correspond to a different strip segmentation.}
		\label{fig:MID_schem}
	\end{figure}

	The MID consists of 72 RPCs arranged in 2 stations of 2 planes each (fig. \ref{fig:MID_schem}). The planes are $\sim$5.5x6.5 m$^2$, with $\sim$1.2x1.2 m$^2$ central hole for the beam pipe and its shielding, and each plane consists of 18 RPCs. A single RPC is $\sim$70x270 cm$^2$ and there are three different shapes, i.e. Long(L), Cut(C) and Short(S). The RPCs are equipped with orthogonal readout strips in order to provide the spatial information along the X and Y directions, i.e. on the plane perpendicular to the beam line, for a total of 21k strips with 1, 2 and 4 cm pitch. The strip pitches increase with the distance from the beam line, in order to provide an almost flat occupancy on the MID surface and to keep the momentum resolution roughly constant all over the detection planes. \\
	To cope with the increased counting rate of Run 3 and to reduce aging effects [10][11], the MID underwent several upgrades which will be described in details in the following sections.
	
	\subsection{MID Resistive Plate Chambers (RPCs)}
	\label{MID_RPC}
	
	MID RPCs are 2mm single gap detectors with 2mm bakelite electrodes, having a $\rho =$ 3x10$^9$ – 1x10$^{10}$ $\Omega$cm. 
	The gas mixture used was composed of 89.7$\%$ C$_2$H$_2$F$_4$, 0.3$\%$ SF$_6$, 10$\%$ \textit{i}-C$_4$H$_{10}$, humidified at 35-40$\%$ RH. During Run 1 and 2 RPC signals were directly sent to front-end discriminators with a threshold of 7 mV, resulting in an operating voltage of about 10.2–10.5 kV at 970 mbar of pressure and 20$^{\circ}$C, in the so-called maxi-avalanche mode (i.e. average charge per hit of 100 pC). 
	
	\subsubsection{RPC status at INFN Torino laboratory}
	\label{RPC_Torino}
	
\begin{figure}[htb]
		\centering
		\includegraphics[width=12.5cm,height=3cm]{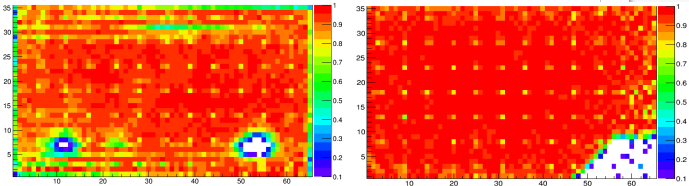}
		\caption{Left panel: Efficiency map at 8500 V. Inefficiency holes are visible and the average efficiency is low all over the RPC surface. Right panel: Efficiency map at 8100 V for one of the Cut type RPCs of the new 2021 production batch.}
		\label{fig:effRPC}
	\end{figure}

	In view of Run 3, a small RPC production was started in 2017 in order to provide some spares to replace gas gaps showing high dark currents or otherwise malfunctioning. Gaps produced before 2019 were however highly unsatisfactory because of inefficiency holes at the HV working point (WP), high currents of the order of several tens of $\mu$A and a general carelessness in the production process (see fig. \ref{fig:effRPC}, left panel). 
	After several interactions with the firm, a new pre-production batch of 3 RPCs was requested at the end of 2019. \\
	The 3 RPCs were tested in 2020 in INFN-To laboratory using cosmic rays and a streamer gas mixture, using the ADULT electronics with 80 mV thresholds, since the tests of the first batch were done with this set-up. 
	The results showed an efficiency higher than 95$\%$ around the WP($\sim$ 8400V for the streamer mode) throughout the entire surface and low currents around 1$\mu$A, i.e. suitable for installetion in ALICE. A new production batch of 30 RPCs was ordered in 2021 but unfortunately there were several delays in the test of this new batch, due to the commissioning of a brand new INFN-Torino laboratory. Four RPCs have been tested so far and a preliminary analisys show that the efficiency plateau is reached at HV even lower than the 2019 batch (WP at $\sim$ 8100V). 
	
	\subsection{New front-end electronics FEERIC and read-out architecture}
	\label{FEERIC}
	
\begin{figure}[htb]
		\centering
		\includegraphics[width=13cm,height=4cm]{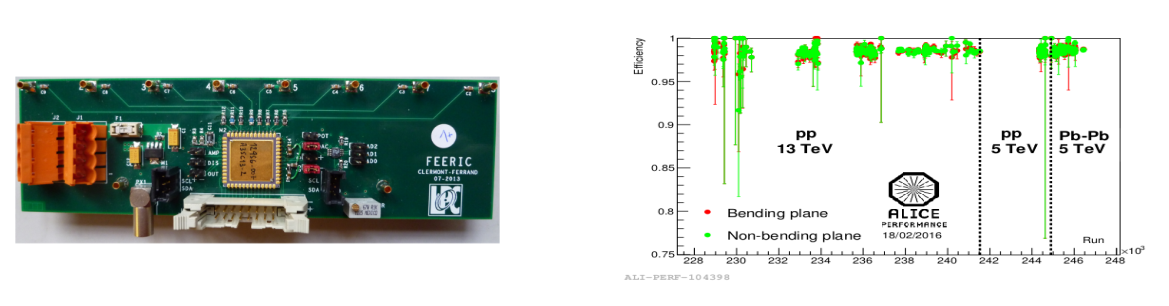}
		\caption{Left: one FEERIC front-end electronics board. Right: efficiency of one RPC equipped with FEERIC as a function of run number, measured during Run 2. For some runs the efficiency is slightly lower due to temporary issues with the electronics.}
		\label{fig:effFEERIC}
	\end{figure}

	The goal after the end of Run 2 was to slow down RPCs aging and improve rate capability. To achieve this and also to cope with the increased counting rate foreseen for Run 3 it was decided to install a  brand new front-end electronics FEERIC [12], which includes a pre-amplification stage, allowing to operate RPC detectors at lower gain. The installation of 2384 FEERIC cards was completed in 2019 and they are now under commissioning with p-p Run 3 data. One RPC was already equipped with FEERIC during Run 2. This RPC had a factor 3-5 less charge released in the gas volume and a lower HV WP. The efficiency was higher than 97$\%$ in both bending and non bending plane 
	as shown in the right panel of fig.\ref{fig:effFEERIC}, where the efficiency is reported as a function of the run number for p-p and Pb–Pb collisions. The RPC equipped with FEERIC showed very satisfactory performances and stability [14]. The thresholds are now adjustable for each single FEERIC card. The Xbee system used to set the thresholds is slow and unstable, so it will be upgraded to WiFi (band 2.4 GHz) during the LHC 2022 winter shutdown. There will be 1 WiFi router per side, connected to the Detector Control System via Ethernet, and 12 WiFi stations per side, connected to FEERIC via I$^2$C protocol. \\
	To cope with the ALICE continuous mode, a new readout electronics was necessary [15]. There are a total of 234 Local cards, up to 16 per VME crate, 16 Regional FPGA-based cards, interfaced with the new Common Readout Unit via 2 GBTx links, and 16 J2-bus between the Local and Regional. 
	The readout cards are under commissioning with p-p data.
	
	\section{MID status at CERN}
	\label{MID_CERN}
	
	Several hardware interventions were necessary during the commissioning without beams, due mostly to gas leakage and HV trips due to several faulty cables and connectors. The average dark current value around the WP, i.e. 9500 V, was 4.59 $\mu$A in 2021 when the MID was turned on for the first time after the end of Run 2. After the interventions in cavern the average dark current of the 72 RPCs decreased to 2.23 $\mu$A at the WP. The entire system was ready when first Run 3 stable beams at top energy arrived on July 5$^{th}$. The task of the new WP HV values and the final adjustment of FEERIC thresholds is ongoing with cosmics and p-p data. 
	
	
	\section{Conclusions}
	\label{Conclu}
	
	During the LS2 the MID has been upgraded with a new front-end and readout electronics to cope with the higher interaction rates of Run 3. FEERIC is in good shape and \textit{in situ} tests proved fully satisfactory. The new gas gap tests are ongoing in INFN-TO laboratory and the results are promising. All the components for the MID data readout chain have successfully been tested and installed in the ALICE cavern. The new RPCs HV working point determination and the FEERIC threshold optimisation are ongoing. In conclusion, regarding the MID status at CERN, all the RPCs are operational and the MID is stable and participating in ALICE data-taking. \\
	
	

\end{document}